# Exact travelling wave solutions for the Penrose-Fife Phase Field Model

**P. O. Mchedlov-Petrosyan[1], L. N. Davydov[2]**

NSC "Kharkov Institute of Physics and Technology"

[1] *E-mail*: peter.mchedlov@free.fr
[2] *E-mail*: ldavydov@kipt.kharkov.ua

The Penrose-Fife Phase Field Model is now a well-established model in the theory of phase transitions. In the course of study of this model both the rigorous mathematical results and approximate solutions were obtained. However, to the best of our knowledge, no exact solutions were given in the literature. In the present paper we give exact travelling wave solutions for this system. While the functional form of the solutions is rather simple, the dependence of solutions on the parameters of the model is quite complicated.

Keywords: phase transition, phase field model, exact solution, travelling wave

## 1. Introduction

The Penrose-Fife Phase Field Model was introduced in [1] (a similar model was independently introduced in [2]) and intensively studied since [3-15]. The standard form of this model is represented by two equations: the first equation results from the energy conservation and the second – describes the evolution of the order parameter. In the first equation the heat flux is presumed to be given by the Fourier law, i.e. to be proportional to the temperature gradient:

$$J = -\sigma T_m \frac{\partial u}{\partial \overline{x}}, \qquad (1.1)$$

where $u$ is the relative deviation of the temperature from the melting temperature $T_m$, $u = (T - T_m)/T_m$, and $\sigma$ is the heat conductivity; in the present work we will presume both $\sigma$ and the specific heat $c$ to be constant. The energy conservation yields the following equation

$$\frac{\partial u}{\partial \overline{t}} - \frac{\partial}{\partial \overline{t}}\left(p\varphi^2 + r\varphi\right) = -\frac{1}{cT_m}\frac{\partial J}{\partial \overline{x}}. \qquad (1.2)$$

The second term in the left-hand side describes the evolution of the order-parameter-dependent ("potential") part of the internal energy. In the Penrose-Fife Phase-Field model internal energy is presumed to be a quadratic (or more generally, concave) function of the nondimensional order parameter $\varphi$, and to have a lower value for the ordered state. The order parameter is presumed to increase gradually from lower value for the disordered (liquid) state to the higher value for the ordered (solid) state, so internal energy should be a decreasing function of $\varphi$. In the Penrose-Fife model the difference of the values of potential part of internal energy between initial and final state,

$$\left(Const - p\varphi^2 - r\varphi\right)\Big|_{liquid} - \left(Const - p\varphi^2 - r\varphi\right)\Big|_{solid} = \overline{L}, \tag{1.3}$$

is the (nondimensional) latent heat $\overline{L} = L/(cT_m)$ released during solidification. Eliminating $J$ from Eqs. (1.1)-(1.2) we get the equation of the standard Penrose-Fife model, with the inverse absolute temperature $1/T$ linearized about the melting temperature (see, e.g.[13]):

$$\frac{\partial u}{\partial \overline{t}} - \frac{\partial}{\partial \overline{t}}\left(p\varphi^2 + r\varphi\right) = K\frac{\partial^2 u}{\partial \overline{x}^2}, \tag{1.4}$$

where $K = \sigma/c$ is the heat diffusivity. For the most applications such a linearization is justified: the absolute temperature $T$ is usually very far from zero for both phases, and it changes in a rather narrow interval between the values for the ordered and disordered phases. The evolution of the order parameter is governed by the following equation

$$\zeta\overline{\theta}\frac{\partial \varphi}{\partial \overline{t}} = \zeta\rho\frac{\partial^2 \varphi}{\partial \overline{x}^2} - \left(\varphi^3 - \delta\varphi^2 - \gamma\varphi - \eta\right) - \left(2p\varphi + r\right)u. \tag{1.5}$$

Slightly different from the standard Penrose-Fife model [1-15] here we have taken a cubic polynomial including the even power terms in the temperature-independent part of the chemical potential, corresponding to the generally asymmetric forth-order polynomial potential in the homogeneous part of the free energy. The coefficient $\overline{\theta}\zeta$ is the characteristic time for the relaxation of the order parameter. The second-derivative term in the right-hand side is due to the input of the inhomogeneities into the free energy; $\zeta\rho$ is usually presumed to be proportional to the square of the capillarity length. It was shown [16-17, 6], that if $\overline{\theta} \sim O(1), \rho \sim O(1)$, and $\zeta$ is a small parameter, both the standard phase-field model and the problem (1.4), (1.5) are asymptotically reduced to a generalized Stefan problem. By proper rescaling of the variables the number of given independent parameters in the system, Eqs. (1.4) and (1.5), is reduced to seven, namely $p, r, \delta, \gamma, \eta, \zeta$ and $\theta = \overline{\theta}K/\rho$; the nondimensional coordinate is $x = \overline{x}/\sqrt{\rho}$, the nondimensional time is $t = \overline{t}K/\rho$.

We assume that far away from the transition region the order parameter $\varphi$ approaches some values $\psi_1$ and $\psi_2$ for the ordered (solid) and for the disordered (liquid) states, respectively:

$$\psi_1 = \varphi\big|_{x=-\infty}; \quad \psi_2 = \varphi\big|_{x=+\infty}; \quad \psi_1 > \psi_2. \tag{1.6}$$

From purely phenomenological point of view there is no reason to attach any particular value neither to $\psi_1$, nor to $\psi_2$. Physically important is the difference between the phases, i.e. the inequality in (1.6). So, the system of equations is

$$\frac{\partial u}{\partial t} - \frac{\partial}{\partial t}\left(p\varphi^2 + r\varphi\right) = \frac{\partial^2 u}{\partial x^2}, \tag{1.7}$$

$$\zeta\theta\frac{\partial \varphi}{\partial t} = \zeta\frac{\partial^2 \varphi}{\partial x^2} - \left(\varphi^3 - \delta\varphi^2 - \gamma\varphi - \eta\right) - \left(2p\varphi + r\right)u, \tag{1.8}$$

and the boundary conditions for the temperature are

$$\left.u\right|_{x=-\infty} = u_1; \quad \left.u\right|_{x=+\infty} = u_2. \tag{1.9}$$

Here $u_1$ and $u_2$ are the temperatures of the ordered (solid) and disordered (liquid) states, respectively, far away from the transition region. We consider the solidification into the supercooled liquid, $u_2 < u_1$; then the latent heat released during solidification is removed via liquid. The temperature of the solid $u_1$ is always not higher than the melting temperature, $u_1 \leq 0$.

In the course of study of the Penrose-Fife Phase-Field model [1-15] both rigorous mathematical results and approximate solutions were obtained. However, to the best of our knowledge, no exact solutions were given in the literature. In the present paper we give exact travelling wave solutions for the system (1.7)-(1.8).

## 2. Travelling wave solution

Looking for the travelling wave solution, $\varphi(z)$, $u(z)$; $z = x - vt$, we get from Eqs. (1.7)-(1.8)

$$v\frac{d}{dz}\left[p\varphi^2 + r\varphi - u\right] = \frac{d^2 u}{dz^2}, \tag{2.1}$$

$$-v\zeta\theta\frac{d\varphi}{dz} = \zeta\frac{d^2\varphi}{dz^2} - \left(\varphi^3 - \delta\varphi^2 - \gamma\varphi - \eta\right) - \left(2p\varphi + r\right)u. \tag{2.2}$$

Let us first consider Eq. (2.1). Integrating this equation once, we get

$$v\left[p\varphi^2 + r\varphi - u + C\right] = \frac{du}{dz}, \tag{2.3}$$

where $C$ is an arbitrary constant. For $z = \pm\infty$ the right-hand side of Eq. (2.3) equals zero, so the left-hand side should be equal to zero as well. Then it follows from the boundary conditions (1.6), (1.9):

$$p\psi_1^2 + r\psi_1 - u_1 + C = 0; \quad p\psi_2^2 + r\psi_2 - u_2 + C = 0. \tag{2.4}$$

Subtracting these equations, we get

$$p\left(\psi_1^2 - \psi_2^2\right) + r\left(\psi_1 - \psi_2\right) = u_1 - u_2. \tag{2.5}$$

On the other hand, in the Penrose-Fife model for the solidification the difference of the values of potential part of internal energy between initial and final state is the (nondimensional) latent heat $\bar{L}$, see Eq. (1.3):

$$\left(Const - p\varphi^2 - r\varphi\right)\Big|_{\varphi=\psi_2} - \left(Const - p\varphi^2 - r\varphi\right)\Big|_{\varphi=\psi_1} = \bar{L} =$$
$$= p\left(\psi_1^2 - \psi_2^2\right) + r\left(\psi_1 - \psi_2\right) \tag{2.6}$$

I.e., Eq. (2.5) means $u_1 - u_2 = \bar{L}$, which, up to notations difference ($c_2 T_m \bar{L} = L$), is exactly the well-known condition for the existence of the constant-velocity travelling-wave solutions of the classic "sharp-boundary" Stefan problem [18-19], see Appendix 1. The existence of the constant-velocity travelling-wave solutions both for the standard phase-field model and Penrose-Fife phase-field model, corresponding (in the small $\zeta$ limit) to this solution of the Stefan problem was proven in [20,9]. However, no exact solutions were found; as we show below, such a solution exists for the Penrose-Fife Phase-Field model.

At this stage we introduce the Ansatz
$$u = \alpha + \beta\varphi, \tag{2.7}$$
i.e. the linear dependence of the temperature on the order parameter, where $\alpha$ and $\beta$ are at present undetermined coefficients. One can easily check that taking higher-order polynomial will not allow balancing nonlinear terms in these equations, see below. It follows immediately from the boundary conditions (1.6), (1.9) that
$$u_1 = \alpha + \beta\psi_1; \quad u_2 = \alpha + \beta\psi_2. \tag{2.8}$$
For the supercooled case which is considered here $u_2 < u_1$, and from Eqs. (2.8) it follows that the inequality
$$\beta > 0 \tag{2.9}$$
should be necessary satisfied (per definition $\psi_1 > \psi_2$). Substitution of the Ansatz (2.7) into Eq. (2.3) for $u$ yields, after some rearrangement
$$vp\left[\varphi^2 - \frac{\beta - r}{p}\varphi + Y\right] = \beta\frac{d\varphi}{dz}, \tag{2.10}$$
where $Y$ is an arbitrary constant. Now, $\psi_1, \psi_2$ should be the roots of the polynomial in the left hand side of Eq. (2.10); i.e.
$$\psi_1 + \psi_2 = \frac{\beta - r}{p}. \tag{2.11}$$
So, if we select the constant $Y$ to be equal to $\psi_1\psi_2$, Eq.(2.10) takes the form
$$\frac{d\varphi}{dz} = \kappa\left(\varphi - \psi_1\right)\left(\varphi - \psi_2\right), \tag{2.12}$$
where

$$\kappa = \frac{pv}{\beta}. \qquad (2.13)$$

We are looking for a monotonically decreasing with $z$ solution (anti-kink). So, it should be $\kappa > 0$. Integrating Eq. (2.12), we obtain

$$\varphi = \frac{\psi_2 + \psi_1 \exp\{-\kappa(\psi_1 - \psi_2)(z + \bar{c})\}}{1 + \exp\{-\kappa(\psi_1 - \psi_2)(z + \bar{c})\}}. \qquad (2.14)$$

If we select $z = 0$ to be the position of the $\max\left(\dfrac{d\varphi}{dz}\right)$, then it should be $\bar{c} = 0$, and Eq. (2.14) may be rewritten as

$$\varphi = \frac{1}{2}(\psi_1 + \psi_2) - \frac{1}{2}(\psi_1 - \psi_2)\tanh\left[\frac{1}{2}\kappa(\psi_1 - \psi_2)z\right]. \qquad (2.15)$$

Substitution of the Ansatz (2.7) for $u$ into Eq. (2.2) yields

$$-v\zeta\theta\frac{d\varphi}{dz} = \zeta\frac{d^2\varphi}{dz^2} - \left[\varphi^3 + (2p\beta - \delta)\varphi^2 + (2p\alpha + r\beta - \gamma)\varphi + (r\alpha - \eta)\right]. \qquad (2.16)$$

Using Eq. (2.12), the derivative $\dfrac{d^2\varphi}{dz^2}$ is easily expressed as a polynomial in $\varphi$; for brevity we introduce notation $X = \psi_1 + \psi_2$; as we already denoted above $\psi_1\psi_2 = Y$ we get

$$\frac{d^2\varphi}{dz^2} = \kappa^2\left[2\varphi^3 - 3X\varphi^2 + (X^2 + 2Y)\varphi - XY\right]. \qquad (2.17)$$

Substituting the latter expression into Eq. (2.16), rearranging and equating to zero coefficients at all powers of $\varphi$ we obtain the following constraints on the parameters

$$\zeta\kappa^2 = \frac{1}{2}, \qquad (2.18)$$

$$v\zeta\theta\kappa = \frac{3}{2}X + 2p\beta - \delta, \qquad (2.19)$$

$$v\zeta\theta\kappa X = \frac{1}{2}(X^2 + 2Y) - (2p\alpha + r\beta - \gamma), \qquad (2.20)$$

$$v\zeta\theta\kappa Y = \frac{1}{2}XY + (r\alpha - \eta). \qquad (2.21)$$

So, if the constraints (2.11), (2.13) and (2.18)-(2.21) are satisfied, $\varphi$ given by Eq. (2.15) and $u$ given by Eq. (2.7) are the solutions of the system (2.1)-(2.2). Taking into account Eqs. (2.8), the solution for $u$, naturally, may be rewritten in the form

$$u = \frac{1}{2}(u_1 + u_2) - \frac{1}{2}(u_1 - u_2)\tanh\left[\frac{1}{2}\kappa(\psi_1 - \psi_2)z\right]. \qquad (2.22)$$

In addition to above mentioned constraints we have also the physical condition

$$u_1 - u_2 = \overline{L}, \qquad (2.23)$$

see Eqs. (2.5)-(2.6), which allows to connect the parameters of the model with the latent heat. Different from the simple form of the solutions, the system of the constraints is rather complicated, see below.

## 3. Parametric dependence of the constant-velocity travelling-wave solution for the Penrose-Fife Phase-Field model

Using Eqs. (2.13) and (2.18) to eliminate $v$ and $\kappa$ from Eqs. (2.19)-(2.21) and taking into account Eq. (2.11), we get the algebraic system for $X$, $Y$, $\alpha$ and $\beta$:

$$\beta = pX + r, \qquad (3.1)$$

$$\frac{\beta\theta}{2p} = \frac{3}{2}X + 2p\beta - \delta, \qquad (3.2)$$

$$\frac{\beta\theta}{2p}X = \frac{1}{2}(X^2 + 2Y) - (2p\alpha + r\beta - \gamma), \qquad (3.3)$$

$$\frac{\beta\theta}{2p}Y = \frac{1}{2}XY + (r\alpha - \eta). \qquad (3.4)$$

From Eq.s (3.1) and (3.2) we easily find $X$ and $\beta$:

$$X = \frac{r(\theta - 4p^2) + 2p\delta}{p(3 + 4p^2 - \theta)}, \qquad (3.5)$$

$$\beta = \frac{3r + 2p\delta}{3 + 4p^2 - \theta} \qquad (3.6)$$

Correspondingly, we get expression for the velocity of the front of phase transition, see Eq. (2.13):

$$v = \frac{3r + 2p\delta}{p\sqrt{2\zeta}(3 + 4p^2 - \theta)} \qquad (3.7)$$

Solving Eqs. (3.3)-(3.4) for $Y$ we obtain the following expression:

$$Y(\theta - 1) = \frac{-\dfrac{r}{2}X^2(\theta - 1) - r^2\left(p + \dfrac{\theta}{2p}\right)X + \gamma r - r^3 - 2p\eta}{(pX + r)}. \qquad (3.8)$$

Correspondingly,

$$\alpha = \frac{1}{2pr} Y \left[ pX(\theta - 1) + \theta r \right] + \frac{\eta}{r}, \tag{3.9}$$

or, substituting Eq. (3.8) for $Y$, we obtain

$$\alpha(\theta - 1) = -\frac{(\theta - 1)pX + \theta r}{2pr(pX + r)} \left[ \frac{r}{2} X^2(\theta - 1) + r^2 \left( p + \frac{\theta}{2p} \right) X - \gamma r + r^3 + 2p\eta \right] + \frac{\eta}{r}(\theta - 1). \tag{3.10}$$

To avoid too lengthy formulae here we consider only the symmetric potential in the homogeneous part of the free energy, i.e. we take $\delta = \eta = 0$. Then expressions (3.5)-(3.8) and (3.10) became

$$X = \frac{r(\theta - 4p^2)}{p(3 + 4p^2 - \theta)}, \tag{3.11}$$

$$\beta = \frac{3r}{3 + 4p^2 - \theta}, \tag{3.12}$$

$$v = \frac{3r}{p\sqrt{2\zeta}(3 + 4p^2 - \theta)}, \tag{3.13}$$

$$Y(\theta - 1) = \frac{-\frac{r}{2}X^2(\theta - 1) - r^2\left(p + \frac{\theta}{2p}\right)X + \gamma r - r^3}{(pX + r)}, \tag{3.14}$$

$$\alpha(\theta - 1) = -\frac{\left[(\theta - 1)X + \frac{\theta r}{p}\right]}{4(pX + r)} \left\{ X\left[(\theta - 1)X + r\left(\frac{\theta}{p} + 2p\right)\right] + 2(r^2 - \gamma) \right\}. \tag{3.15}$$

Here we presume $\theta \neq 1$; the special "degenerate" case $\theta = 1$ is considered in Appendix 2. Then the values of the order parameter $\psi_1$ and $\psi_2$ for the ordered and disordered phases, respectively, are

$$\psi_{1,2} = \frac{1}{2}X \pm \sqrt{\frac{1}{4}X^2 - Y}, \tag{3.16}$$

where $X$ and $Y$ are given by Eqs. (3.11) and (3.14). The values $\psi_{1,2}$ should be real, which imposes a condition

$$X^2 > 4Y \tag{3.17}$$

on the values of the parameters of the model. Also, quite similar to the case of the constant-velocity travelling wave for the sharp boundary problem, see Appendix 1, in addition to the system of constraints (3.1)-(3.4) we always have the *physical* condition $u_1 - u_2 = \overline{L}$, see Eq. (2.23). This imposes an additional constraint on the parameters of the model:

$$\beta^2 \left( X^2 - 4Y \right) = \overline{L}^2. \tag{3.18}$$

Evidently, if Eq. (3.18) is satisfied, inequality (3.17) is fulfilled automatically. Eliminating from the latter equation $\beta$ and $Y$ but keeping $X$ (given by Eq. (3.11)) to have not too bulk formula we obtain

$$\frac{pX+r}{\theta-1}\left[\left(pX+3r\right)X^2\left(\theta-1\right)+4r^2\left(p+\frac{\theta}{2p}\right)X+4r^3-4\gamma r\right]=\overline{L}^2. \tag{3.19}$$

This equation yields an additional constraint imposed on the four parameters of the model $p$, $r$, $\gamma$, and $\theta$. I.e., the model is characterized by any three parameters from the above four, $\zeta$, and the latent heat $\overline{L}$. The coefficients $\alpha$ and $\beta$ in the Ansatz, Eq. (2.7), are given by Eqs. (3.15) and (3.12), respectively. So the values of the reduced temperature, which correspond to the bulk ordered and disordered phases, are

$$u_1 = \alpha + \beta\psi_1; \quad u_2 = \alpha + \beta\psi_2. \tag{3.20}$$

## 4. Discussion

The most essential difference of the Penrose-Fife Phase Field model from the traditional one is the start from the entropy as thermodynamic potential, and internal energy and inverse absolute temperature as conjugate variables [1], while the traditional model starts from the free energy, and temperature and entropy as the conjugate variables. Additionally, the internal energy is presumed to be a nonlinear (concave) function of the order parameter. Then the entropy functional becomes non-decreasing along the solution path. The nonlinearity of the internal energy as function of the order parameter appears also to be crucial for the existence of the *exact* constant-velocity travelling wave solution.

Remarkably, such a solution has a direct analogue in the well known sharp boundary problem [18,19], see also Appendix 1. Indeed, the exact constant-velocity travelling wave solution of the sharp-boundary problem exists for special temperatures of the solid and liquid phases only, see Eqs. (5.13)-(5.14) of the Appendix 1. Similarly, if the parameters of the model, Eqs. (1.7)-(1.8), are given, the values of the reduced temperature for the bulk solid and liquid phases are also fixed, see Eqs. (3.20).

Our initial hypothesis was that there should be a polynomial link between the temperature and order parameter; it appeared, however, that taking higher-order polynomial will not allow balancing nonlinear terms in equations (1.7)-(1.8), so our Ansatz was reduced to the linear link between these variables.

The functional form of the solutions, Eqs. (2.15) and (2.22), is rather simple. Quite opposite, the dependence of $\psi_{1,2}$, i.e. the values of the order parameter far away from the transition region, of the steepness of the transition front, etc., on the parameters of the model is quite complicated. Still, some general observations are possible. Remarkably, $\psi_{1,2}$, $\alpha$, $\beta$ and, consequently, $u_{1,2}$ depend on $p$, $r$ and $\delta$, $\gamma$, $\eta$, i.e. on the parameters in the expressions for internal and free energy and on the ratio of the relaxation times for the order parameter and temperature fields $\theta$. On the other hand, the velocity $v$ and the steepness $\kappa(\psi_1 - \psi_2)$ of the front (see Eqs. (3.13) and (2.18)) depend additionally on $\zeta$, i.e. on the scale of inhomogeneity.

To make the formulae somewhat more transparent starting from Eq. (3.11) we considered only the symmetric potential in the homogeneous part of the free energy, i.e. we take $\delta = \eta = 0$. An interesting distinction arises in connection with the values of the parameter $\theta$. It is evident from Eq. (3.14) that there is a special, "degenerate" case $\theta = 1$, which is considered in some detail in Appendix 2. It means physically that the characteristic times for the evolution of the order parameter, $(\bar{\theta}\zeta)$, and for the heat transfer, $\zeta\rho/K$, on the spatial scale $\sqrt{\zeta\rho}$ of the transition domain are *exactly* equal. It is worth mentioning that in some modifications of the model $\theta$ was set equal to unity from the very beginning; however for our exact solution it appears to be a rather special case. Indeed, for this case only one constraint is imposed on the stationary values $\psi_1$ and $\psi_2$ of the order parameter, $\psi_1 + \psi_2 = X$, where $X$ is given by Eq. (3.11). On the other hand, for this case from Eq. (3.14) arises additional a constraint (6.3), which allows to eliminate $r$.

So, while for the general case $\theta \neq 1$ there are finally five parameters and an additional physical constraint (3.19), for the special case $\theta = 1$ the solution depends on three independent parameters. The parametric "degrees of freedom" of the model allow, in principle, an "inverse" approach: one may consider, e.g. $u_{1,2}$ as given values and adjust the parameters of the model correspondingly.

## Appendix 1

Here we reiterate for convenience the well known result for the sharp boundary problem between two phases ("Stefan Problem"), see e.g. [18,19]. Let us consider the plane boundary at $x = \Gamma(t)$, separating solid and liquid phases. $T\big|_{x=\Gamma(t)} = T_m$, where $T_m$ is the melting temperature. The temperature of the solid phase far away from the boundary is $T_1$, the temperature of the liquid phase far away from the boundary is $T_2$. The temperature of the solid is lower than (or equal to) the melting temperature, $T_1 \leq T_m$; we consider the case when $T_2 < T_m$. The evolution of the temperature field in both domains is governed by the standard heat conduction equations,

$$\frac{\partial T}{\partial t} = \kappa_1 \frac{\partial^2 T}{\partial x^2}, \quad -\infty < x < \Gamma(t), \tag{5.1}$$

$$\frac{\partial T}{\partial t} = \kappa_2 \frac{\partial^2 T}{\partial x^2}, \quad \Gamma(t) < x < +\infty, \tag{5.2}$$

where $\kappa_i$ are the thermal diffusivities, i.e. thermal conductivities, divided by specific heats. The condition at the advancing solidification boundary is

$$L\frac{d\Gamma}{dt} = c_1\kappa_1 \left.\frac{\partial T}{\partial x}\right|_{x=\Gamma-0} - c_2\kappa_2 \left.\frac{\partial T}{\partial x}\right|_{x=\Gamma+0}, \tag{5.3}$$

where $L$ is the latent heat, and $c_i, i=1,2$, specific heat for solid and melt, respectively. This condition simply means that the latent heat released during solidification should be moved away from the boundary.

For arbitrary $T_1$, $T_2$ the above formulated problem has the well known self-similar solution in terms of the $erf$ (error) – functions

$$T = \frac{(T_m - T_1)\operatorname{erf}\left(\dfrac{x}{2\sqrt{\kappa_1 t}}\right) + T_1\operatorname{erf}\left(\dfrac{\lambda}{2\sqrt{\kappa_1}}\right) + T_m}{1 + \operatorname{erf}\left(\dfrac{\lambda}{2\sqrt{\kappa_1}}\right)}, \quad -\infty < x < \Gamma(t), \tag{5.4}$$

$$T = \frac{-(T_m - T_2)\operatorname{erf}\left(\dfrac{x}{2\sqrt{\kappa_2 t}}\right) - T_2\operatorname{erf}\left(\dfrac{\lambda}{2\sqrt{\kappa_2}}\right) + T_m}{1 - \operatorname{erf}\left(\dfrac{\lambda}{2\sqrt{\kappa_2}}\right)}, \quad \Gamma(t) < x < +\infty \tag{5.5}$$

Here

$$\lambda = \frac{\Gamma(t)}{2\sqrt{t}}. \tag{5.6}$$

is the self-similar coordinate of the solidification front. Substitution of Eqs. (5.4) and (5.5) into boundary condition (5.3) yields the following equation for finding $\lambda$:

$$\lambda L = \sqrt{\frac{\kappa_1}{\pi}} \frac{c_1(T_m - T_1)\exp\left(-\dfrac{\lambda^2}{\kappa_1}\right)}{1 + \operatorname{erf}\left(\dfrac{\lambda}{\sqrt{\kappa_1}}\right)} + \sqrt{\frac{\kappa_2}{\pi}} \frac{c_2(T_m - T_2)\exp\left(-\dfrac{\lambda^2}{\kappa_2}\right)}{1 - \operatorname{erf}\left(\dfrac{\lambda}{\sqrt{\kappa_2}}\right)} \tag{5.7}$$

However, for the special values of $T_1$, $T_2$ there is a constant-velocity travelling wave solution. Introducing the travelling wave coordinate, $z = x - vt$, where

$v = \dfrac{d\Gamma}{dt} = \text{const}$ , we get instead of Eqs. (5.1), (5.2) ordinary differential equations

$$-v\frac{dT}{dz} = \kappa_1 \frac{d^2T}{dz^2}, \quad -\infty < z < 0, \tag{5.8}$$

$$-v\frac{dT}{dz} = \kappa_2 \frac{d^2T}{dz^2}, \quad 0 < z < +\infty, \tag{5.9}$$

and instead of Eq. (5.3)

$$Lv = c_1 \kappa_1 \left.\frac{dT}{dz}\right|_{z=-0} - c_2 \kappa_2 \left.\frac{dT}{dz}\right|_{z=+0}. \tag{5.10}$$

These equations have solutions

$$T = \text{const}, \quad -\infty < z < 0, \tag{5.11}$$

$$T = T_2 + \left(T_m - T_2\right)\exp\left(-\frac{v}{\kappa_2}z\right), \quad 0 < z < +\infty. \tag{5.12}$$

Evidently, the solution (5.11) could match the boundary conditions if, and only if

$$T_1 = T_m. \tag{5.13}$$

Substitution of Eq. (5.12) into (5.10) yields finally

$$L = c_2\left(T_m - T_2\right). \tag{5.14}$$

I.e., such the solution exists only for the special value of supercooling; the velocity of the solidification front remains undetermined [18,19]. It is worth mentioning that for these special values of $T_1$ and $T_2$ the self-similar solution (5.5) does not exist. Indeed, setting $T_1 = T_m$ in Eq. (5.7) we get

$$\lambda\sqrt{\frac{\pi}{\kappa_2}}\exp\left(\frac{\lambda^2}{\kappa_2}\right)\left(1 - \text{erf}\left(\frac{\lambda}{\sqrt{\kappa_2}}\right)\right) = \frac{c_2\left(T_m - T_2\right)}{L}. \tag{5.15}$$

If additionally Eq. (5.14) is fulfilled, the right-hand side of Eq. (5.15) is equal to one; then this equation is satisfied in the limit $\lambda \to \infty$ only [21].

The constant-velocity planar-boundary solution was used mainly for pedagogical purposes and to model special processes (e.g. directional solidification) where the velocity of the front is prescribed. However the three-dimensional analogue of this solution with the constant-velocity non-planar boundary, either of the form of paraboloid of revolution, or the paraboloidal cylinder (the famous Ivantzov solutions [22]) are widely used in modeling dendrites, etc. Remarkably, the analogues planar-boundary, constant-velocity solution exists for the melting problem as well, i.e. for $v < 0$. Then the solutions are

$$T = \text{const}, \quad 0 < z < \infty, \tag{5.16}$$

$$T = T_1 - (T_1 - T_m)\exp\left(-\frac{v}{\kappa_1}z\right), \quad -\infty < z < 0. \tag{5.17}$$

Evidently, the solution (5.16) could match the boundary conditions if, and only if $T_2 = T_m$. Substitution of Eq. (5.17) into (5.10) yields finally

$$L = c_1(T_1 - T_m). \tag{5.18}$$

I.e., for melting with constant velocity the melt should be kept at the melting temperature, and the solid should be overheated according to Eq. (5.18).

## Appendix 2

It is evident from Eq. (3.14) that there is a special, "degenerate" case $\theta = 1$. It means physically that the characteristic times for the evolution of the order parameter, $\overline{\theta}\zeta$, and for the heat transfer, $\zeta\rho/K$, on the spatial scale $\sqrt{\zeta\rho}$ of the transition domain are *exactly* equal. It is worth mentioning that in some modifications of the model $\theta$ was set equal to unity from the very beginning; however for our exact solution it appears to be a rather special case. Indeed, for this case one constraint only is imposed on the stationary values $\psi_1$ and $\psi_2$ of the order parameter, $\psi_1 + \psi_2 = X$, where $X$ is given by Eq. (3.11):

$$X = \frac{r(1 - 4p^2)}{2p(1 + 2p^2)}. \tag{6.1}$$

The unknown $Y = \psi_1\psi_2$ disappears from Eq.(3.14); but instead, this equation becomes an additional necessary constraint imposed on the parameters of the model, i.e. $p$, $r$, and $\gamma$:

$$r\left(p + \frac{1}{2p}\right)X - \gamma + r^2 = 0. \tag{6.2}$$

Substitution of Eq. (6.1) for $X$ in the latter equation allows to express $r$ as function of $p$ and $\gamma$

$$r^2 = 4p^2\gamma. \tag{6.3}$$

Similarly, $\alpha$ disappears from Eq. (3.15) if $\theta = 1$; the remaining constraint is, naturally, again Eq. (6.2). However, we still have the link between $\alpha$ and $Y$ given by Eq. (3.9), which for $\eta = 0$ and $\theta = 1$ simplifies essentially:

$$Y = 2p\alpha. \tag{6.4}$$

If $\theta = 1$, the expression (3.12) for the coefficient $\beta$ also simplifies to

$$\beta = \frac{3r}{2(1 + 2p^2)}. \tag{6.5}$$

Substituting Eq. (6.3) for $r$ into Eqs. (6.1) and (6.5), we get $X$ and $\beta$ as functions of $p$ and $\gamma$,

$$X = \frac{\sqrt{\gamma}\left(1 - 4p^2\right)}{\left(1 + 2p^2\right)}, \tag{6.6}$$

$$\beta = \frac{3p\sqrt{\gamma}}{\left(1 + 2p^2\right)}. \tag{6.7}$$

On the other hand, there is the physical constraint (2.23), which yields an additional equation for $\psi_1$ and $\psi_2$:

$$\psi_1 - \psi_2 = \frac{\bar{L}}{\beta}. \tag{6.8}$$

Using Eqs. (6.6)-(6.8) we get the system to determine $\psi_1$ and $\psi_2$:

$$\psi_1 + \psi_2 = \frac{\sqrt{\gamma}\left(1 - 4p^2\right)}{\left(1 + 2p^2\right)}, \tag{6.9}$$

$$\psi_1 - \psi_2 = \bar{L}\frac{\left(1 + 2p^2\right)}{3p\sqrt{\gamma}}. \tag{6.10}$$

I.e., the stationary values of the order parameter at $\mp\infty$ are

$$\psi_{1,2} = \frac{1}{2}\left[\frac{\sqrt{\gamma}\left(1 - 4p^2\right)}{\left(1 + 2p^2\right)} \pm \bar{L}\frac{\left(1 + 2p^2\right)}{3p\sqrt{\gamma}}\right]. \tag{6.11}$$

Consequently, $Y = \psi_1\psi_2$ and $\alpha$ (see Eq. (6.4)) become

$$Y = \frac{1}{4}\left[\frac{\gamma\left(1 - 4p^2\right)^2}{\left(1 + 2p^2\right)^2} - \bar{L}^2\frac{\left(1 + 2p^2\right)^2}{9p^2\gamma}\right], \tag{6.12}$$

$$\alpha = \frac{1}{2p}Y = \frac{1}{8p}\left[\frac{\gamma\left(1 - 4p^2\right)^2}{\left(1 + 2p^2\right)^2} - \bar{L}^2\frac{\left(1 + 2p^2\right)^2}{9p^2\gamma}\right]. \tag{6.13}$$

Finally, using Eqs. (2.8), (6.7), (6.11), and (6.13) we get the expressions for $u_{1,2}$; the velocity of the transition front is

$$v = \frac{3}{1 + 2p^2}\sqrt{\frac{\gamma}{2\zeta}}. \tag{6.14}$$